# Modeling Decoherence and Decoherence-Free Subspaces

Jeffrey Satinover[1]

*Department of Physics, Yale University, New Haven, Connecticut*

(Comments: 22 pages, 18 figures) A simplified Bogoliubov transform reduces a fully-interacting many-fermion spin-1/2 system-plus-environment to a more tractable many-to-one variant. The transform additionally yields exact solutions for bosonic multi-particle interactions sans the approximation introduced by using discrete time steps to deal with quantum parallelism. The decohering effect of relatively general finite environments is therewith modeled and compared to the decohering effect of an infinite environmental "bath." The anti-symmetric singlet state formed by two maximally-entangled two-state particles is shown to be inherently decoherence-free. As a quantum bit ("qubit") it is thus potentially superior to any single-particle state.

is accomplished by operating in a controlled way (using quantum "gates") upon a system of quantum particles in superposed states deemed "qubits" when working with two-state systems such as spin ½ particles in states such as:

$$|\psi\rangle = \alpha|1\rangle + \beta|0\rangle \qquad (1)$$

Effective quantum information processing requires that the evolution of qubit states take place coherently, i.e., preserving the relative (phase) relationship between component kets that existed at initialization of the state. In this way, a single operation can be carried out in parallel on two different states per qubit, that is, on $2^N$ $N$-bit states encoded by $N$ qubits. By contrast, two-state classical information processing involves operations upon one state per bit, that is a single $N$-bit state encoded by $N$ bits.

The major obstacle to the construction of QUIP-capable devices is the fact that uncontrolled interaction among the system particles—or between the system particles and the environment—will rapidly cause the initial phase relations between component kets in the qubit to vary. In theory, for a small and perfectly-known set of environmental couplings, this variation is periodic and can be tracked and accounted for. In practice, a mere handful of uncorrelated extraneous interactions will cause the phase variation quickly to become multi-periodic (i.e., decoherent) with rare and transient revivals of relative coherence. The mean fidelity (overlap between the initial and time-developed state vectors) is on the order of $2^{-N/2}$, where $N$ is the number of environmental particles. (E.g., ten environmental particles will cause a system particle in the state (1) to become, on average, ~97% decohered.)

To date, the two most widely-explored techniques to combat decoherence are isolation of the system from the environment and regularly-timed error-correction. A third, newer approach is conceptually different from both of these: There exist a subset of quantum states that, in principal, are inherently resistant to decoherence. These are certain entan-gled *multi-particle* states whose internal anti-symmetry causes environmental interactions to cancel out. (They are, in effect, multi-particle eigenstates of the interaction hamiltonian.) Such *decoherence-free subspaces* (DFS's) of the system particles' total Hilbert space have therefore been proposed as an alternative computational basis.

---
[1] jeffrey.satinover@yale.edu    http://pantheon.yale.edu/~satinovr

Previous papers on decoherence have suggested numerical methods for modeling decoherence for a single spin-½ particle system in a multi-spin-½ environment [1,2] using random matrices; have explicated a number of key mathematical arguments for the existence of DFS's [3,4]; and have experimentally verified that simple physical examples can be created and perform as expected [5].

It is generally considered necessary to simulate time-developed spin-spin interaction using some kind of discrete approximation (e.g., time-steps, and sliding or randomized sequencing in the fashion of synchronous or asynchronous cellular automata) to deal with quantum parallelism in serial fashion [6-8].

This paper demonstrates how the Hamiltonian for any fully-interacting set of fermions or bosons (or both) may instead be rewritten using a simplified Bogoliubov transformation as a Hamiltonian describing an arbitrary number of quasi-particles identified as the "system," with the remainder identified as "environment," and for which all environmental quasi-particles interact with all system particles, but without internal interactions among either system or environment. This reduced degree of parallelism may be handled serially, either numerically or analytically, without approximation. For those situations that are of greatest usefulness with respect to a DFS, only the coupling effects among system particles need be known and quantified.

As the environment grows larger, the numerical simulation approaches the analytic equations that govern decoherence of a system in an infinite environmental "bath." The simplified form of the environment-system interaction Hamiltonian can thus serve as a realistic proxy for a fully-interacting system plus environment, both with respect to modeling decoherence and in understanding the conditions that allow for a DFS in the system.

When the system is not a single two-state particle, but a two-particle four-state system, the dynamics of decoherence can be significantly different if the system particles are entangled. Examples of such entangled two particle states are the four "Bell States":

$$\frac{|00\rangle+|11\rangle}{\sqrt{2}}, \frac{|01\rangle+|10\rangle}{\sqrt{2}}, \frac{|00\rangle-|11\rangle}{\sqrt{2}}, \frac{|01\rangle-|10\rangle}{\sqrt{2}} \qquad (2)$$

From both theoretical calculations and numerical simulations, an important relationship emerges between the various entangled states of the two-particle system and the overall degree of decoherence caused by the environment. In particular, the fourth state in (2), a maximally-entangled, fully anti-symmetric state, is in theory 100% resistant to an environment acting in *any* basis, or in any mixture of bases.

Decoherence of a single spin-½ particle in a state of the form (1) in the $\sigma_z$ basis (the "system") may be realistically modeled by examining the time-evolution of the off-diagonal elements of the 2 X 2 reduced density matrix for the system particle. It interacts with an environment consisting of an arbitrary number of spin-½ particles, all acting in arbitrary bases $\sigma_n = a_1\sigma_x + a_2\sigma_y + a_3\sigma_z$, all interacting with each other as well, and with an arbitrary set of coupling constants. For any meaningful number of particles, however, the interaction Hamiltonian would be too complex to work with. A number of reductions can be made.

First, if an environmental spin acts in a basis $\sigma_x$ orthogonal to the system spin $\sigma_z$ it will cause the system spin to precess around the x-axis with period τ. The state (1) with $\alpha = \beta$, for example, will therefore at some time τ/2 have evolved as follows:

$$\frac{|0\rangle + |1\rangle}{\sqrt{2}} \xrightarrow{\frac{\tau}{2}} \frac{|1\rangle + |0\rangle}{\sqrt{2}} \qquad (3)$$

Indeed, under action in the $\sigma_x$ basis, the *relative* phase angle θ between the two $\sigma_z$ kets will remain the same at all times, and only a phase factor common to both, hence ignorable, will be introduced. The same is true for action in the $\sigma_y$ basis. In short, *only effects in the system spin basis can cause system decoherence.* Hence to simulate decoherence, we may reduce our environment to one composed exclusively of $\sigma_z$ acting spins.[*]

Second, and more importantly, we may reduce a set of spins each with arbitrary couplings to every other, to a set of environmental (quasi-) spins that all interact with only a single arbitrarily selected system spin. We may select more than one spin as system and do likewise. These mathematical transformations have three benefits. (A) Modeling: We may accurately model decoherence of a single system particle using an environment that is exponentially simplified: Instead of $2^N$ interactions (all-to-all), we need only simulate $bN$ (where $b = 1$ for all-to-one, $b > 1$ for all-to-some). (B) Theory: DFS's result from certain internal symmetries in the *system*. The presence of such symmetries may be evident only in a special basis. By transforming the system basis to one in which intra-system interactions are eliminated (between quasi-particles), these symmetries can be uncovered. Otherwise they are hidden by the asymmetrical time-evolution of the more natural particle kets that form the "laboratory" basis. (C) Physical Applications: We may work backwards and devise operators that act in strange and seemingly useless ways in the particle basis but that do what we wish in the non-interacting quasi-particle basis.

We first show how a set of $N$ mutually-interacting spins may be transformed into $N$-1 quasi-spins all interacting with the remaining one. Consider a set of spin ½ particles, in some fixed, random arrangement in space, all acting in the pure $\sigma_z$ basis, all interacting with each other. The interaction Hamiltonian may be written as:

$$\hbar \sum_{i=0}^{N} \sum_{j=0}^{N} \omega_{ij} \left( \sigma_i \otimes \sigma_j \prod I_k \right) (1 - \delta_{ij}) \qquad (4)$$

$$= \hbar \sum_{i=0}^{N} \sum_{j=0}^{N} \omega_{ij} \sigma_i \sigma_j (1 - \delta_{ij}) \qquad (5)$$

The factor $(1 - \delta_{ij})$ excludes self-interaction; there are no restrictions on the interaction strength $\omega_{ij}$; I is the 2 X 2 identity matrix for the $k^{th}$ Hilbert space. Setting $N = 1$, we may rewrite $H_{int}$ in terms of a Hermitian coefficient matrix:

---

[*] For QUIP, however, action in the $\sigma_x$ or $\sigma_y$ basis will cause problems other than decoherence. For example, many quantum computation gates deliberately flip selected spins at synchronized time-steps as part of a desired algorithm. An environment that flips spins arbitrarily will introduce errors.

$$(\sigma_0 \quad \sigma_1 \quad \cdots \quad \sigma_n) \begin{bmatrix} \omega_{00} & \omega_{01} & \cdots & \omega_{0N} \\ \omega_{01}^* & \omega_{11} & \cdots & \omega_{1N} \\ \vdots & \vdots & \ddots & \vdots \\ \omega_{0N}^* & \omega_{1N}^* & \vdots & \omega_{NN} \end{bmatrix} \begin{pmatrix} \sigma_0^+ \\ \sigma_1^+ \\ \vdots \\ \sigma_N^+ \end{pmatrix} \equiv \boldsymbol{\sigma^+ H \sigma} \quad (6)$$

There exists an $N \times N$ unitary matrix **U**, its elements $(\mathbf{U}_{ij}) = u_{ij}$ depending upon the $(\mathbf{H}_{ij}) = \omega_{ij}$, such that:

$$\mathbf{U^+ \; H \; U \; = \; H_D}$$

where $\mathbf{H_D}$ is diagonal, with real elements $\Omega_{ii}$:

$$\begin{bmatrix} \Omega_{00} & 0 & \cdots & 0 \\ 0 & \Omega_{11} & \cdots & 0 \\ \vdots & \vdots & \ddots & \vdots \\ 0 & 0 & \cdots & \Omega_{NN} \end{bmatrix} \quad (7)$$

Therefore, $\mathbf{H = U\, H_D\, U^+}$, so that $H_{int} = \boldsymbol{\sigma^+ H \sigma} = \boldsymbol{\sigma^+} (\mathbf{U H_D U^+}) \boldsymbol{\sigma} = (\boldsymbol{\sigma^+} \mathbf{U}) \mathbf{H_D} (\mathbf{U^B} \boldsymbol{\sigma})$
where:

$$(\mathbf{U^+ \sigma}) = \begin{bmatrix} u_{00} & u_{10}^* & \cdots & u_{N0}^* \\ u_{10} & u_{11} & \cdots & u_{1N} \\ \vdots & \vdots & \ddots & \vdots \\ u_{N0} & u_{N1} & \vdots & u_{NN} \end{bmatrix} \begin{pmatrix} \sigma_0^+ \\ \sigma_1^+ \\ \vdots \\ \sigma_N^+ \end{pmatrix} = \begin{pmatrix} \sigma_0'^+ \\ \sigma_1'^+ \\ \vdots \\ \sigma_N'^+ \end{pmatrix} \quad (8)$$

$$= \boldsymbol{\sigma'} = \begin{pmatrix} u_{00}\sigma_0^+ + u_{10}\sigma_1^+ + \cdots u_{N0}\sigma_N^+ \\ u_{10}^*\sigma_0^+ + u_{11}\sigma_0^+ + \cdots + u_{1N}\sigma_0^+ \\ \cdots \\ u_{N0}^*\sigma_N^+ + u_{N1}^*\sigma_N^+ + \cdots u_{NN}\sigma_N^+ \end{pmatrix} \quad (9)$$

and similarly for $\boldsymbol{\sigma'^+}$. In this "rotated" basis, each particle contributes to a basis of quasi-particles represented by a set of quasispins $\sigma_j'$. The z-orientation remains unaffected as the quasiparticles are composed of mixtures of spins from different Hilbert spaces, but still acting in the z-basis.

The rotated interaction Hamiltonian,

$$H' = (\sigma'_0 \quad \sigma'_1 \quad \cdots \quad \sigma'_n) \begin{bmatrix} \Omega_{00} & 0 & \cdots & 0 \\ 0 & \Omega_{11} & \cdots & 0 \\ \vdots & \vdots & \ddots & \vdots \\ 0 & 0 & \cdots & \Omega_{NN} \end{bmatrix} \begin{pmatrix} \sigma_0'^+ \\ \sigma_1'^+ \\ \vdots \\ \sigma_N'^+ \end{pmatrix} = \Omega_{00}\sigma_0' \sigma_0^{+'} + \Omega_{11}\sigma_1' \sigma_1^{+'} + \cdots + \Omega_{NN}\sigma_N' \sigma_N^{+'} \quad (10)$$

$$= \sum_{i=0}^{N} \Omega_{ii} \sigma_i' \sigma_i^{+'} \quad (11)$$

is that *for N non-interacting (quasi-)spins*. To obtain a Hamiltonian for $N-1$ spins interacting only with the first, we perform the same procedure on just the lower right $(N-1)\times(N-1)$ portion of the original coefficient matrix. The new coefficient matrix :

$$\begin{bmatrix} \omega_{00} & \omega_{01} & \cdots & \omega_{0N} \\ \omega^*_{01} & \Omega_{11} & \cdots & 0 \\ \vdots & \vdots & \ddots & \vdots \\ \omega^*_{0N} & 0 & \cdots & \Omega_{NN} \end{bmatrix} \quad (12)$$

yields the desired many-to-one Hamiltonian:

$$(\omega'_{01}\sigma_0\sigma'_1 + \omega^{*'}_{01}\sigma'_1\sigma_0 + \ldots + \omega'_{0N}\sigma_0\sigma'_N + \omega^{*'}_{0N}\sigma'_N\sigma_0) + (\omega_{00}\sigma_0\sigma_0 + \Omega_{11}\sigma'_1\sigma'_1 + \ldots + \Omega_{NN}\sigma'_N\sigma'_N) \quad (13)$$

with

$$\frac{\hbar}{2}\sum_{i=0}^{N}\omega_{ij}\sigma_0\sigma^+_i = \frac{\hbar}{2}\sum_{i=0}^{N}\omega_{ij}\sigma_0\sigma_i \quad (14)$$

The first term in (13) represents a system of (quasi-)spins *all of which are interacting with the first spin* ($\sigma_0$; *not* a quasi-particle!) but not with each other. The second term represents a system of spins all of which are interacting with themselves only. Decoherence of the system spin is due to the effect on $\sigma_0$ of all the environmental spins—taken as a whole. The net effect on $\sigma_0$ of all the environmental *quasi*-spins—taken as a whole—is, of course, exactly the same, since $\sigma_0$ is the same in both representations. Thus, to model decoherence of a system spin in a full-interacting environment of spins acting in arbitrary bases we need only simulate a many-to-one interaction Hamiltonian with all spins acting in a common basis.

Suppose a normalized initial state for the $\sigma_i$:

$$|\psi_{\text{sys}}(0)\rangle = \alpha_0|0\rangle + \beta_0|1\rangle \quad (15)$$

$$|\varphi_{\text{env},i}(0)\rangle = \alpha_i|0\rangle + \beta_i|1\rangle \quad (16)$$

with $|\alpha_i|^2 + |\beta_i|^2 = 1$, $i = 0, 1, \ldots N$. The various $\alpha_i \in \mathbb{C}$ ( and so, too, the respective $\beta_i$ ) are unrelated to one another, and are located at random in the complex square defined by $\pm 1$ and $\pm i$. Initially, the complete state vector is:

$$(\alpha_0|0\rangle + \beta_0|1\rangle)_{\text{sys}} \otimes \prod_{i=1}^{N} \otimes (\alpha_i|0\rangle + \beta_i|1\rangle)_{\text{env}} \quad (17)$$

The reduced density matrix for $\sigma_0$ is:

$$\begin{pmatrix} |\alpha_0|^2 & \alpha_0\beta_0^* \\ \beta_0\alpha_0^* & |\beta_0|^2 \end{pmatrix} \quad (18)$$

The non-zero values of the off-diagonal elements indicate the presence of the initial superposition.

The interaction Hamiltonian is:

$$\sum_{i=1}^{N}(\omega_{0i}\sigma_0\sigma_i + \omega^*_{0i}\sigma_i\sigma_0) = \sum_{i=1}^{N}(\omega_{0i} + \omega^*_{0i})\sigma_i\sigma_0 \tag{19}$$

and the propagator it gives rise to is of the form:

$$e^{iH_{\text{int}}t} = e^{i\sum_{i=1}^{N}\lambda_i^{(\alpha)}\omega_i t} \otimes e^{-i\sum_{i=1}^{N}\lambda_i^{(\beta)}\omega_i t} \tag{20}$$

It has two "branches" (separated by the tensor product sign) that operate independently with the |0⟩ and |1⟩ components of the system. In a manner of speaking, one represents the perturbation of the environment induced by the |0⟩ component of the system (labeled by its coefficient α) and the other by the |1⟩ component of the system (labeled by β). The value of each pair ($\lambda^{(\alpha)}_i$, $\lambda^{(\beta)}_i$) will be $(+1,-1)$ or $(-1,+1)$ depending on whether it is the $\sigma_i$ eigenvalue corresponding to the |0⟩ or to the |1⟩ component of the $i^{\text{th}}$ environmental spin. (The ± contributed by the system spin state has already been separated out as the negative in the second exponential.) These will also branch into two. Each pair of branches will act in its own space, one branch multiplied by the coefficient $\alpha_i$, the other by the coefficient $\beta_i$. (Thus, if there were only two particles, there would be four separate exponentials in the overall wavefunction—two tensored by two; if three particles, six—two tensored by two tensored by two.)

The time developed state vector is no longer separable:

$$\left|\Psi_{sys+env}(t)\right\rangle = \alpha_0|0\rangle\prod_{i=1}^{N}\otimes\left(\alpha_i e^{i\omega_i t}|0\rangle + \beta_i e^{-i\omega_i t}|1\rangle\right) + \beta_0|1\rangle\prod_{i=1}^{N}\otimes\left(\alpha_i e^{-i\omega_i t}|0\rangle + \beta_i e^{i\omega_i t}|1\rangle\right) \tag{21}$$

We construct a full density matrix for this state. For the factors multiplying $|0\rangle\langle 1|_{\text{sys}}$, we obtain:

$$\alpha_0\beta_0^*|0\rangle\langle 1|_{sys} \otimes \prod_{i=1}^{N}\otimes\left(|\alpha_i|^2 e^{2i\omega_i t}|0\rangle\langle 0| + \alpha_i\beta_i^*|0\rangle\langle 1| + \beta_i\alpha_i^*|1\rangle\langle 0| + |\beta_i|^2 e^{-2i\omega_i t}|1\rangle\langle 1|\right) \tag{22}$$

When we trace out the environmental factors to get the reduced density matrix, we get:

$$\alpha_0\beta_0^*|0\rangle\langle 1|_{sys} \otimes \prod_{i=1}^{N}\otimes\left(|\alpha_i|^2 e^{2i\omega_i t}\langle 0||0\rangle + \alpha_i\beta_i^*\langle 1||0\rangle + \beta_i\alpha_i^*\langle 0||1\rangle + |\beta_i|^2 e^{-2i\omega_i t}\langle 1||1\rangle\right) \tag{23}$$

$$= \alpha_0\beta_0^*|0\rangle\langle 1|_{sys} \otimes \prod_{i=1}^{N}\left(|\alpha_i|^2 e^{2i\omega_i t} |\beta_i|^2 e^{-2i\omega_i t}\right)$$

Or, in terms of cosines and sines:

$$\alpha_0\beta_0^*|0\rangle\langle 1|_{sys}\prod_{i=1}^{N}\left[\cos(2\omega_i t) + \left(|\alpha_i|^2 - |\beta_i|^2\right)i\sin(2\omega_i t)\right] \tag{24}$$

Similarly, for the other off-diagonal element:

$$\beta_0\alpha_0^*|1\rangle\langle 0|_{sys}\prod_{i=1}^{N}\left[\cos(2\omega_i t) - \left(|\alpha_i|^2 - |\beta_i|^2\right)i\sin(2\omega_i t)\right] \tag{25}$$

Conclusion: *The off-diagonal terms in the reduced density matrix rapidly tend toward zero for any significant number of environmental particles.* The reason is that the factors in (25) are numbers that vary at unrelated rates (determined by the $\omega_i$ between unrelated extrema (determined by the $\alpha_i$, $\beta_i$). Furthermore, the absolute value of these extrema are always less than 1 for $\alpha_i \neq \beta_i$ (when they equal 1) and the different numbers attain their extrema and only for $\omega_i t/\pi = 2n$, n = 0, 1, 2,… i.e., at unrelated times. The product of many cosines (sines) of unrelated frequency is *almost always a very small number*.

No matter how many unrelated cosines enter into it, however, the product of a finite number of cosines is nonetheless strictly speaking still periodic, with multiple sub-periods ("multiperiodic"). The off-diagonal elements of the reduced density matrix can therefore always re-attain any arbitrary proportion of their initial value within some finite time. The time-developed *average* value, however, is on the order of $2^{-N/2}$ and this average value is reached extremely quickly as the following simulations show.

To simulate the above, a *mathematica* program was developed that models any number of system and environment spins (up to computational capacity) acting in any basis with arbitrary (or otherwise distributed) coupling strengths and acting in any bases. The simplifications discussed above allow the simulation to be carried out using only $z$-basis spins and a many-to-one coupling Hamiltonian.

Example 1. Decoherence and the Size of the Environment

(a) We look at the degree of fractional coherence in the system spin at time $t = 1$ as a function of the number of spins in the environment. The first spin (1 on the horizontal axis) is the system spin.

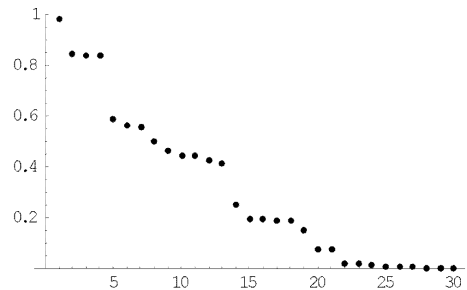

FIG. 1: Horizontal Axis: Number of Spins; Vertical Axis: Proportion of Coherence at $t = 1$.

(b) A second example of the same, with the environmental couplings and spin-states re-randomized.

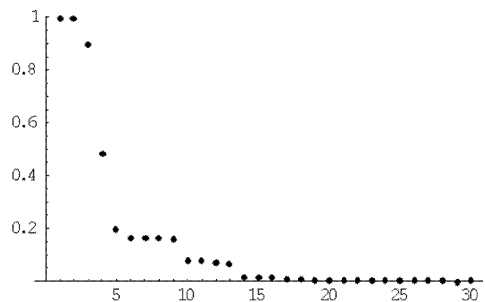

FIG. 2: Horizontal Axis: Number of Spins; Vertical Axis: Proportion of Coherence at $t = 1$.

Example 2. Decoherence and Time

(a)  We look at thirty spins and examine the decoherence that occurs between $t = 0$ and $t = 1$.

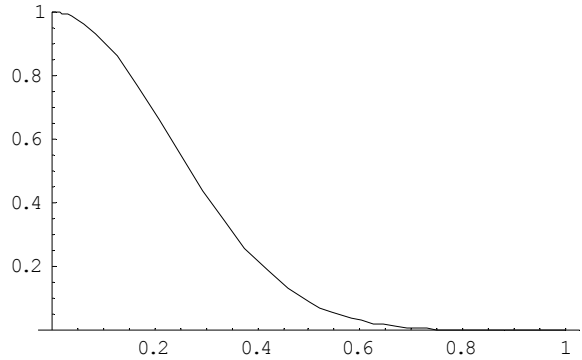

FIG. 3: Horizontal Axis: Time; Vertical Axis: Proportion of Coherence for Thirty Spins.

(b)  Decoherence over a time-span ten times longer (the spins have been re-randomized)

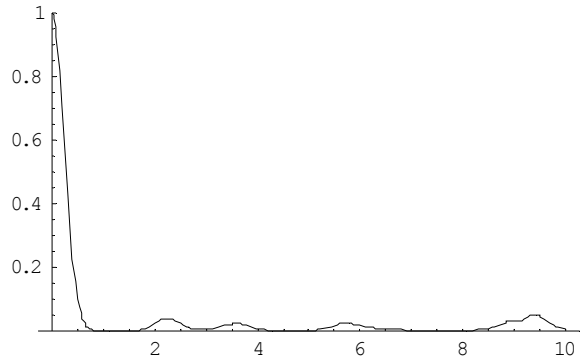

FIG. 4: Horizontal Axis: Time; Vertical Axis: Proportion of Coherence for Thirty Spins.

Note the occasional forays into re-coherence at long time intervals. A finite number of spins each with a finite-precision coupling strength will generate a coefficient with a finite frequency. Eventually, such as system will return to a fully coherent state and repeat the same pattern. The time-scale on which this occurs is indicated by the fact that the average degree of coherence (vertical axis) approaches $\sim (1/\sqrt{2})^n$. With 100 spins, the average remaining coherence is therefore $\sim 9 \times 10^{-16}$. A typical numerical result for 100 randomized spins using the method above at $t = 1$ had a value of $1.03194 \times 10^{-16}$.

Example 3. Decoherence and the Size of the Environment over Time

(a)  We set the coupling strengths for 24 spins and plot the time development of the remaining coherence in the system spin as we progressively add one additional environmental spin. The time ranges from 0 to 30.

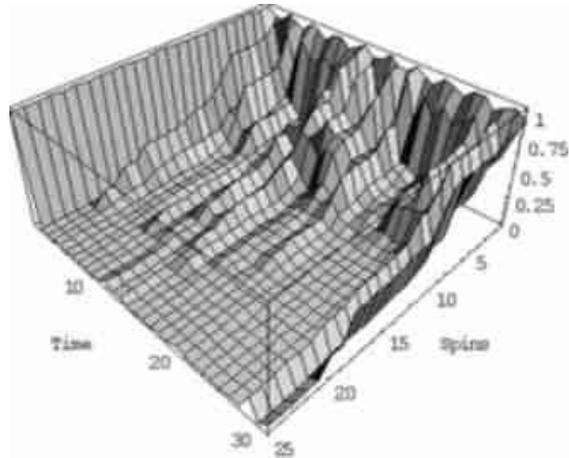

FIG. 5: Left Horizontal Axis: Time; Right Horizontal Axis: Number of Spins; Vertical Axis: Proportion of Coherence.

(b)   The same, with up to 50 spins and a time range from 0 to 100.

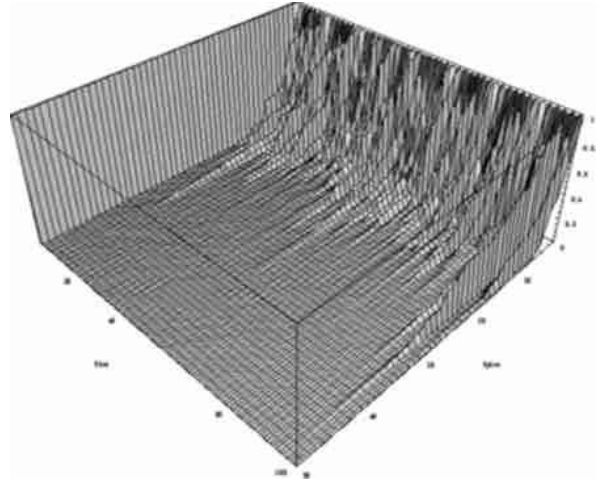

FIG. 6: Left Horizontal Axis: Time; Right Horizontal Axis: Number of Spins; Vertical Axis: Proportion of Coherence.

As the number of spins tends to infinity, the mean coherence approaches zero and the decoherence rate infinite. Even the revival that is apparent in Fig. 6. for up to 30 spins at $t = 96$ (in this particular randomization) is for all practical purposes absent for more than 50 spins. Furthermore, the "wall" at the upper left represents the essentially instant plummeting of coherence from 1 to just about zero, for any number of spins more than, say, ten. Nonetheless, when a physically realistic environment is statistically huge the decoherence is *not* effectively instantaneous and total.

In any finite volume-shell surrounding a target system of finite size there can be only a finite number of environmental spins and therefore a finite net contribution to decoherence from that shell. Each additional finite "shell" contributes a new volume that grows as the square of the distance. But the (dipole) coupling strength falls off as the inverse cube of the distance. Thus, the effective coupling strengths to the system spin for a uniform infinite environment forms a (natural log) distribution. The shells with an ever *larger* number of spins contribute an ever *smaller* effect. The distribution of coupling con-stants thus has a mean of zero ($\omega_i \to 0$ for a spin shell at $r \to \infty$), and tails to the left and right ($\omega_i \to \pm\infty$ at a hypothetical zero).

The integral of a $1/r$ distribution between 1 and $\infty$ does not converge (1 corresponds to zero distance—the normalized radius of the system itself). But it grows ever more slowly. Thus, an environment larger enough than the system to be treated in the statistical limit thermodynamically, but small enough to be physically realistic will have so many spins that it may be handled mathematically as a continuum, but *not as an infinite-sized one.* Furthermore, in most physically realistic situations, the environment has some kind of actual boundary, even if this is unknown.

Mathematically we may express these considerations in the aggregate by foregoing a uniform density of environmental spins with some unknown boundaries. We choose a Gaussian distribution of coupling strengths, the integral of which, of course, does converge. (This is a quite conservative replacement: If the spin density itself as a function of distance were Gaussian, the integrated coupling strength to the system would converge far more rapidly.) The infinite product of every spin in such an environment is in the form of a product of the sum of exponentials (which we rewrote before as sines and cosines), now with $N = \infty$:

$$\alpha_0 \beta_0^* |0\rangle\langle 1| \prod_{i=1}^{\infty} \left( e^{2i\omega_i t} |\alpha_i|^2 + e^{-2i\omega_i t} |\beta_i|^2 \right) \tag{26}$$

No matter what basis the environmental spins are expressed in, the effect on the reduced density matrix for the system spin (in a given basis) will obviously be the same. (The "Schmidt Decomposition" theorem proves this formally.) Without loss of generality, we may therefore write (26) more simply as:

$$\alpha_0 \beta_0^* |0\rangle\langle 1| \prod_{i=1}^{\infty} e^{2i\omega_i t} = \alpha_0 \beta_0^* |0\rangle\langle 1| e^{\sum_{i=1}^{\infty} 2i\omega_i t} \tag{27}$$

which is the same expression we would obtain if all the environmental spins were initially in either the $|0\rangle$ or $|1\rangle$ eigenstate. We now pass to the integral: Instead of ordering the spins by the discrete index $i$ and running it from 1 to infinity, we order the environment as a continuum in coupling strength, from most strongly negative $(-\infty)$ through weakest (0) through most strongly positive $(+\infty)$. The index $i$ that individually identified each spin is now most simply replaced by the value of the coupling strength itself, ω. We assume a variance of 0.2, yielding the distribution shown in Fig. 7. (We loosen this restriction on the variance later.)

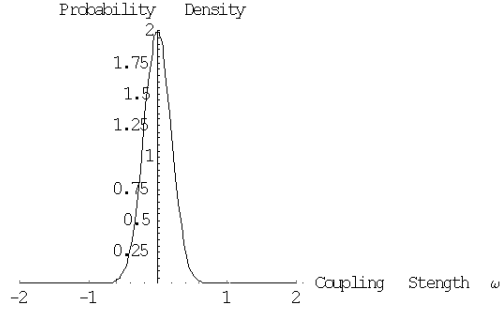

FIG. 7. Distribution of coupling strengths to a finite system at the center of an infinite continuous spherical "sea" of identical spins. The spin density falls off with distance such that the distribution is Gaussian.

The vast preponderance of spin regions thus have coupling strengths ω′ that fall between −1 and +1.

To compute the total decoherence effect on the system spin, we must first multiply each possible value for a coupling strength by its fractional incidence in the entire population, normalized to 1, and integrate over all ω′:

$$\frac{1}{\sqrt{4\pi\lambda}} \int_0^\infty e^{-\frac{\omega'^2}{4\lambda}} |\omega'| d\omega' = \sqrt{\frac{\lambda}{\pi}} \tag{28}$$

The exponential in (27) thus becomes $e^{2i\sqrt{\frac{\lambda}{\pi}}t}$.

Now, from the self-integrating characteristic of the natural logarithm, we re-express (28) as an integral involving itself, and allow the *variance* in the spin coupling strength to assume a Gaussian distribution of possibilities:

$$e^{2i\sqrt{\frac{\lambda}{\pi}}t} = \frac{1}{\sqrt{4\pi\mu}} \frac{2it}{\sqrt{\pi}} \int_{-\infty}^\infty e^{2i\sqrt{\frac{\lambda}{\pi}}t} e^{-\frac{\lambda}{4\pi\mu}} d(\sqrt{\frac{\lambda}{\pi}}) \tag{29}$$

The evaluated integral applied to the reduced density matrix leads to the key result for the magnitude of the off-diagonal element as a function of time:

$$\alpha_0 \beta_0^* e^{-4\mu t^2} |0\rangle\langle 1| \tag{30}$$

where $\mu$ represents the transformed variance. With an infinite number of spins, and a Gaussian distribution of coupling strengths treated as a continuous environment, the off-diagonal terms vanish as $e^{-\mu t^2}$ as shown in Fig. 8, that is to say, exponentially in the range of coupling strengths and exponentially in the square of the time.

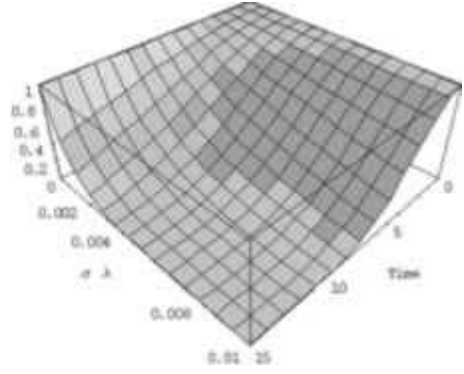

FIG. 8. Decoherence as a function of time and as of the degree of variability in a surrounding infinite environment.

That any environment is bound to generate at least *some* minimum spread in one or the other follows from the energy-time uncertainty relations. Decoherence is apparently a dynamic effect that cannot ever be avoided entirely by manipulations of the environment alone.

There is another way of arriving at the same result. The origin of (28) is the environment, denoted particle-by-particle. Particle 1 acts on the system, then particle 2, and so on. The derivation in (19) to (30) preserves this physical picture; the product form arises from the way we slice the pie.

But the trace operation on the environment requires that one *sum* precisely the same terms. In the infinite, continuous limit, with the weighting provided by the distribution and adding some notational shortcuts, we obtain:

$$\frac{\alpha_0 \beta_0^*}{\sqrt{4\pi\lambda}} \int_{-\infty}^{\infty} (e^{i\omega t}) |0\rangle\langle 1| (e^{-i\omega t})^* e^{\frac{\omega^2}{4\lambda}} d\omega = \frac{\alpha_0 \beta_0^*}{\sqrt{4\pi\lambda}} \int_{-\infty}^{\infty} U(t) |0\rangle\langle 1| U^+(t) e^{\frac{\omega^2}{4\lambda}} d\omega \qquad (31)$$

which evaluates to:

$$\alpha_0 \beta_0 |0\rangle\langle 1| e^{-4\lambda t^2} \qquad (32)$$

as before, with λ in place of μ and where $U(t)$ is an operator on the system. The so-called "operator sum representation" is therefore:

$$\rho = \frac{1}{\sqrt{4\pi\lambda}} \int_{-\infty}^{\infty} (e^{i\omega t}) |\Psi_{sys+env}\rangle\langle\Psi_{sys+env}| (e^{-i\omega t})^* e^{\frac{\omega^2}{4\lambda}} d\omega = \frac{1}{\sqrt{4\pi\lambda}} \int_{-\infty}^{\infty} (e^{i\omega t}) |\psi_{sys}\rangle\langle\psi_{sys}| (e^{-i\omega t})^* |\omega\rangle\langle\omega| e^{\frac{\omega^2}{4\lambda}} d\omega \qquad (33)$$

The second expression in (33) may be written as shown, with |ω⟩⟨ω| separated out, only if the environment is initially not entangled with the system. Otherwise, this step is skipped and one proceeds directly to the trace operation:

$$\rho_{sys} = \frac{1}{\sqrt{4\pi\lambda}} Tr_{env}[\int_{-\infty}^{\infty} (e^{i\omega t}) |\Psi_{sys+env}\rangle\langle\Psi_{sys+env}| (e^{-i\omega t})^* e^{\frac{\omega^2}{4\lambda}} d\omega] = \frac{1}{\sqrt{4\pi\lambda}} \int_{-\infty}^{\infty} \langle\omega| (e^{i\omega t}) |\psi_{sys}\rangle\langle\psi_{sys}| (e^{-i\omega t})^* |\omega\rangle e^{\frac{\omega^2}{4\lambda}} d\omega \qquad (34)$$

so that

$$\frac{1}{\sqrt{4\pi\lambda}} \int_{-\infty}^{\infty} (e^{i\omega t}) |\psi_{sys}\rangle\langle\psi_{sys}| (e^{-i\omega t})^* \langle\omega|\omega\rangle e^{\frac{\omega^2}{4\lambda}} d\omega = \frac{1}{\sqrt{4\pi\lambda}} \int_{-\infty}^{\infty} (e^{i\omega t}) |\psi_{sys}\rangle\langle\psi_{sys}| (e^{-i\omega t})^* e^{\frac{\omega^2}{4\lambda}} d\omega = \sum_{i,j=0}^{1} \int_{-\infty}^{\infty} (e^{i\omega t}) |i\rangle\langle j| (e^{-i\omega t})^* e^{\frac{\omega^2}{4\lambda}} d\omega \qquad (35)$$

We may now compare some typical finite results with those in the limit of an infinite environment. Using the same random Gaussian distribution of spins with λ = 0.2, Figs. 9, 10 and 11 show numerical simulations of decoherence with 5, 10 and 50 environmental spins, respectively, over ten time units. Fig. 12 shows the analytic solution for an infinite number of environmental spins treated as above as a Gaussian-distributed continuum:

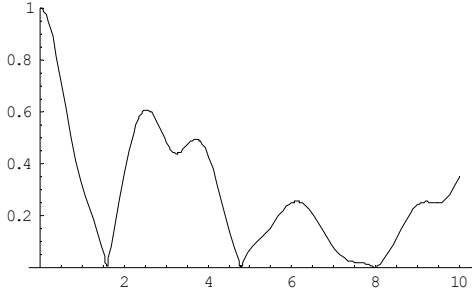

FIG. 9: Decoherence due to 5 spins over $0 \leqslant t \leqslant 10$.

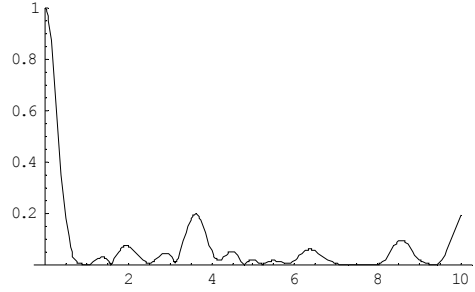

FIG. 10: Decoherence due to 10 spins over $0 \leqslant t \leqslant 10$.

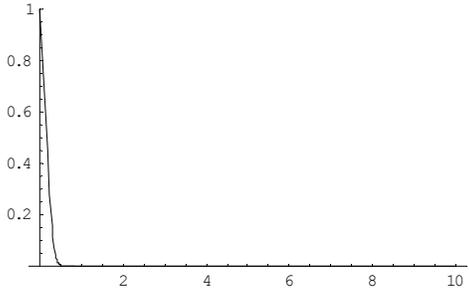

FIG. 11: Decoherence due to 50 spins over $0 \leqslant t \leqslant 10$

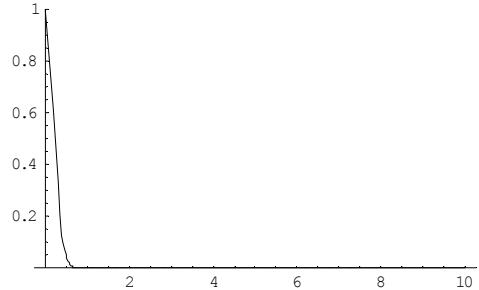

FIG. 12: Decoherence due to an infinite, continuous "sea" of spins with a Gaussian distribution of coupling strengths over $0 \leqslant t \leqslant 10$

Figs. 13-16 contrast coherence over time between an infinite Gaussian environment and an *averaged* range of finite ones. Ten separate randomized simulations were performed, each using the same parameters for the $\alpha_i$ and $\beta_i$ The $\alpha_i$ were selected uniformly at random from within the interval [0, 1] and the $\beta_i$ calculated such that the sum of their squares = 1; the $\omega_i$ were selected at random from a normal distribution over $(-\infty, +\infty)$ with mean = 0 and λ = 0.2. Decoherence was then computed for each set of parameters, at each of twenty time intervals of 0.25, for each of twenty different numbers of environmental spins (0 spins, 10 spins, 20 spins, … , 200 spins). These results were then averaged.

Fig. 13 shows the superimposed results of ten different randomized runs, each for an environment of fifty spins and using the above parameters. The time runs from 0 to 5 units, each unit consisting of 4 ticks. Fig. 14 displays the averaged results for the same set of simulations.

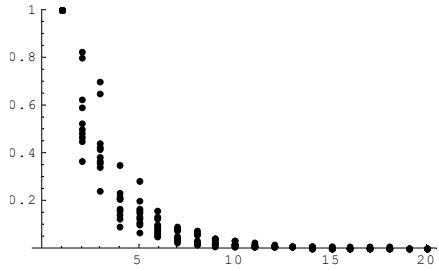
FIG. 13: Decoherence due to fifty spins, ten simulations superimposed.

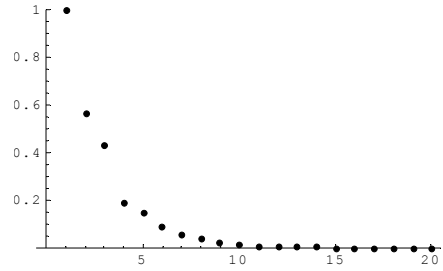
FIG. 14: Decoherence due to fifty spins, ten simulations averaged.

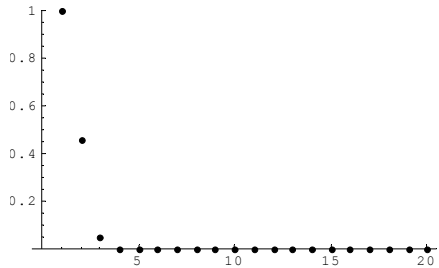
FIG. 15: Decoherence due to an infinite environment.

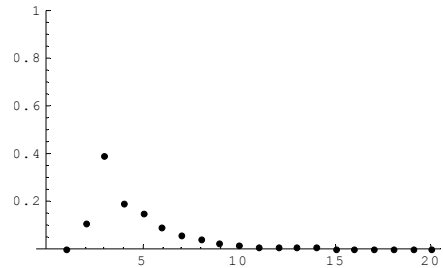
FIG. 16: Difference between the decoherence due to the average of ten simulations of twenty spins and that due to an infinite environment

Fig. 15 shows the decoherence due to an infinite environment over the same duration, and Fig. 16 the difference between it and a ten-run mean for twenty spins. We see that after some time, only the *average* of many different finite environments enforces a monotonic decrease of coherence. We conclude that, so long as the integrated effect of any *particular* infinite environment converges, it too will also display coherence revivals of any arbitrary degree in some finite time. Only those environments that ensure precisely the right out-of-phase relations among all the infinite spins—hand designed or arising, we imagine, by happenstance—could circumvent such revivals. Since any real environment is in fact discrete; and since no physical value is ever infinitely precise (if for no other reason than quantum uncertainty itself), we conclude that revivals of coherence, however rare, are as inevitable in principle as decoherence itself.

This conclusion is of more than academic interest. Decoherence may be characterized as a *dispersion of information* from the system into the environment. For most practical purposes such dispersion is permanent. Hence, in the practical approximation, the time evolution of the decoherent Hamiltonian is non-unitary, i.e., irreversible. But if a revival occurs, then that same information has been recovered, at least briefly. Or, to express it more precisely, decoherence does *not* imply—as does, for example, "collapse of the wave function"—that information has been *permanently destroyed.* (Because such revivals are rare, they are mischaracterized as "spurious," and decoherence is thereby offered as an explanation for measurement—i.e., state-vector "collapse." This is clearly not so.)

Furthermore, the fact that information is not truly *lost* via decoherence—merely dispersed into the environment from whence it may be recollected—raises the at-least theoretical possibility that *it can be contained locally in the first place*. Just as friction and dissipation are classically irresistible thermodynamic "forces" (on a statistical basis) yet can under certain conditions be evaded via judiciously applied quantum phenomena (e.g., superconductivity), so likewise might the even more irresistible-seeming "force" of decoherence.

Surprisingly, however, the maximum decoherence of a *two*-particle state in a large environment of spins is not inevitably ~100% as in the single-state systems simulated above, but can range anywhere downward from ~100% to exactly zero. This maximum depends on the relative values of the coefficients α, β, γ and δ for the two-particle state vector:

$$(a_0|0\rangle + a_1|1\rangle) \otimes (b_0|0\rangle + b_1|1\rangle) = \alpha|0\rangle|0\rangle + \beta|0\rangle|1\rangle + \gamma|1\rangle|0\rangle + \delta|1\rangle|1\rangle \quad (36)$$
$$= \alpha|00\rangle + \beta|01\rangle + \gamma|10\rangle + \delta|11\rangle$$

with $\sqrt{\alpha^2 + \beta^2 + \gamma^2 + \delta^2} = 1$.

To compare the one- and two-particle situations, we first re-express as a topographic map the four elements of a simulated *single*-particle reduced density matrix (Fig. 17), with the height of each of its four quadrants proportional to the remaining coherence for |0⟩⟨0|, |0⟩⟨1|, |0⟩⟨1| and |1⟩⟨1|. We embed the plot as four-element (dim = 2) sub-space in a sixteen-element (dim = 4) space, since it is to the larger space that we will shortly compare it, and observe it over time. The on-diagonal elements persist; the off-diagonal elements quickly vanish.

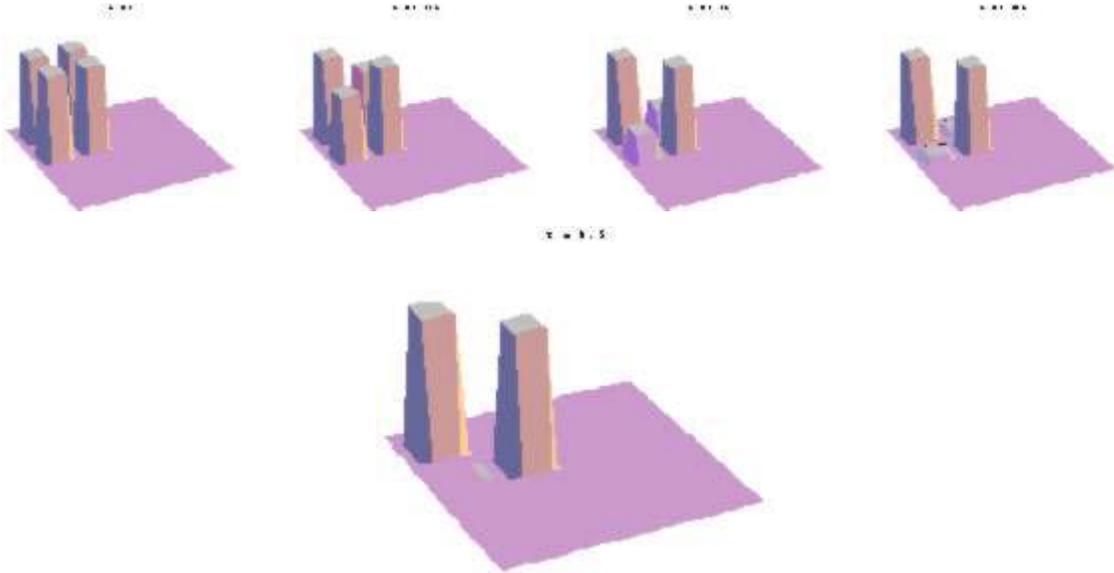

Fig. 17. Density matrix simulation over time for a single-particle system spin in the $\frac{1}{\sqrt{2}}(|0\rangle + |1\rangle)$ state. The environment consists of ten spins, with arbitrary coupling strengths to the system, all acting in the z-basis and all in the state |0⟩. The diagonal |0⟩⟨0| and |1⟩⟨1| quadrants are at the top left and bottom right of each plot; the off-diagonal |0⟩⟨1| and |1⟩⟨0| are catty-corner.

. Next, we choose the two-particle system state:

$$|\Psi\rangle = \tfrac{1}{2}(|00\rangle + |01\rangle + |10\rangle + |11\rangle) \quad (37)$$

The environment is the same. The reduced density matrix for *one* environmental spin acting upon this system is:

$$\begin{pmatrix} \frac{1}{4} & \frac{1}{4}e^{2i\omega t} & \frac{1}{4}e^{2i\omega t} & \frac{1}{4}e^{2i\omega t} \\ \frac{1}{4}e^{-2i\omega t} & \frac{1}{4} & \frac{1}{4} & \frac{1}{4}e^{2i\omega t} \\ \frac{1}{4}e^{-2i\omega t} & \frac{1}{4} & \frac{1}{4} & \frac{1}{4}e^{2i\omega t} \\ \frac{1}{4}e^{-2i\omega t} & \frac{1}{4}e^{-2i\omega t} & \frac{1}{4}e^{-2i\omega t} & \frac{1}{4} \end{pmatrix} \quad (38)$$

the measurable component of which is:

$$\begin{pmatrix} \frac{1}{4} & \frac{1}{4}\cos(2\omega t) & \frac{1}{4}\cos(2\omega t) & \frac{1}{4}\cos(2\omega t) \\ \frac{1}{4}\cos(2\omega t) & \frac{1}{4} & \frac{1}{4} & \frac{1}{4}\cos(2\omega t) \\ \frac{1}{4}\cos(2\omega t) & \frac{1}{4} & \frac{1}{4} & \frac{1}{4}\cos(2\omega t) \\ \frac{1}{4}\cos(2\omega t) & \frac{1}{4}\cos(2\omega t) & \frac{1}{4}\cos(2\omega t) & \frac{1}{4} \end{pmatrix} \quad (39)$$

For an environment of ten spins of random coupling strength, each element in the reduced density matrix will therefore be composed of a product of cosines, e.g. ¼ cos($2t\omega_1$)($2t\omega_2$)…($2t\omega_{10}$)[2]. Fig. 18 illustrates the time development of decoherence in the sixteen subspaces.

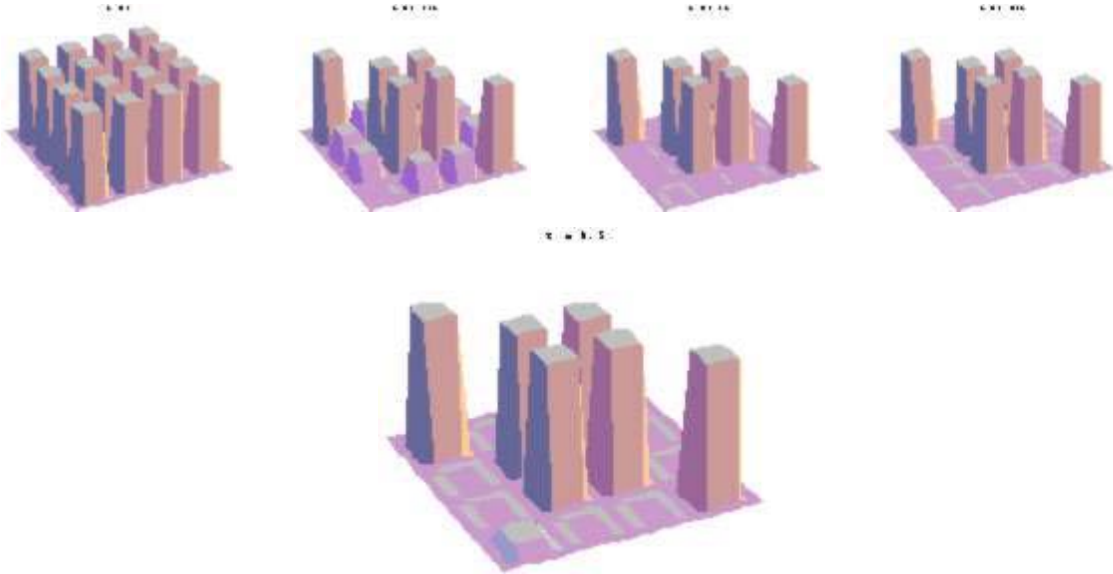

Fig. 18. Density matrix simulation over time for a two-particle system spin in the ½ ( |00⟩ + |01⟩ + |10⟩ + |11⟩ ) state.

The elements most far off-diagonal vanish most quickly. More importantly, we see that within the four dimensional Hilbert space of the two-particle system, *there exists a 2 X 2 (dim = 2) subspace in which no decoherence whatever occurs*. Two of the four elements in the reduced density matrix of this sub-space belong to the diagonal of the larger space itself; that these elements do not vanish is of no interest: They are generated by eigen-states. But that the remaining two elements persist is both surprising and potentially useful: They are generated by superpositions that occur within certain subspaces, but evidently not all.

---

[2] Recall that when we used a more general environment composed of arbitrary superpositions acting in the *z*-basis, each factor was of the form $ab^*\cos(\omega_i t) + (\alpha_i^2 - \beta_i^2) i \sin(\omega_i t)$. When $\alpha_i = \beta_i$, the imaginary term vanishes. In all cases, the real (measurable) component consists only of the cosine term.

Note that these persisting off-diagonal elements are located dead center in the density matrix. In a two-particle universe, the density matrix for a single particle identified in isolation from all others can only occupy a corner on the larger diagonal. The density matrix for the other particle must occupy the other corner, also on the diagonal.

The sub-space for this decoherence-free "particle" is therefore neither one particle nor the other, yet somehow both—but not wholly both either, since the true two-particle density matrix is not 2 X 2 (dim = 2) but 4 X 4 (dim = 4). So what is this subspace/particle?

The two off-diagonal elements in the center of the density matrix are those that arise from the system components $|01\rangle\langle10|$ and $|10\rangle\langle01|$ after tracing out the environment. These specific components of $\alpha|00\rangle + \beta|01\rangle + \gamma|10\rangle + \delta|11\rangle$ *are resistant to decoherence* (or, "this dim = 2 subspace $\{|01\rangle,|10\rangle\}$ within the dim = 4 Hilbert space spanned by $\{|00\rangle,|01\rangle,|10\rangle,|11\rangle\}$" is decoherence-free).

For a two-particle system interacting with a certain kind of environment, this "Decoherence-Free Subspace" consists of *that part of a two-particle, separable state which, when extracted from the whole is no longer separable but entangled*, e.g.:

$$\tfrac{1}{\sqrt{2}}\left(|01\rangle+|10\rangle\right) \tag{40}$$

Its protected status arises from an internal symmetry: Whereas $|0\rangle$ and $|1\rangle$ are affected by the environment in equal but opposite ways, $|01\rangle$ and $|10\rangle$ are affected by the environment in *identical* ways. However, the state defined by (40) is 100% decoherence free *only vis a vis an environment acting purely in the z-basis*. If the environment were acting in the *x*-basis, for example, then a rotation of the basis of representation to the action basis will convert Eq 1.42 to:

$$\tfrac{1}{\sqrt{2}}\left(|00\rangle+|11\rangle\right)_x \tag{41}$$

Like its single-particle analog $\tfrac{1}{\sqrt{2}}\left(|0\rangle+|1\rangle\right)$, the state represented by (41) is fully decohered. The time-developed state *in the x-basis* rapidly approaches a 50:50 statistical mixture of $|00\rangle_x$ and $|11\rangle_x$ states, sans any remaining superposition, except for rare, minor revivals. Transforming back to the *z*-basis, we have the same 50:50 statistical mixture rewritten as:

$$50\%|00\rangle_x = \tfrac{1}{2}\left(|00\rangle_z+|01\rangle_z+|10\rangle_z+|11\rangle_z\right) \tag{42}$$

and

$$50\%|11\rangle_x = \tfrac{1}{2}\left(|00\rangle_z-|01\rangle_z-|10\rangle_z+|11\rangle_z\right) \tag{43}$$

As these states represent different particles in the overall ensemble, there is no single state vector for everything, as there was at the start, and we therefore cannot simply can-cel the middle terms. We form two separate reduced density matrices and add them. The result is:

$$\begin{pmatrix} \frac{1}{4} & 0 & 0 & \frac{1}{4} \\ 0 & \frac{1}{4} & \frac{1}{4} & 0 \\ 0 & \frac{1}{4} & \frac{1}{4} & 0 \\ \frac{1}{4} & 0 & 0 & \frac{1}{4} \end{pmatrix} \quad (44)$$

Compare this to the reduced density matrix for the initial state:

$$\begin{pmatrix} 0 & 0 & 0 & 0 \\ 0 & \frac{1}{2} & \frac{1}{2} & 0 \\ 0 & \frac{1}{2} & \frac{1}{2} & 0 \\ 0 & 0 & 0 & 0 \end{pmatrix} \quad (45)$$

We see at a glance that the original terms, both on and off-diagonal have shrunk by 50%. They have been partially dispersed, as it were, into four new states that were not present at all at the start. 50% of the originally coherent superposition remains present. More formally, we define the fidelity as:

$$\left(Tr\sqrt{\sqrt{\rho_0}\rho(t)\sqrt{\rho_0}}\right)^{\frac{1}{2}} = \left(Tr\sqrt{\sqrt{|\psi_0\rangle\langle\psi_0|}|\psi(t)\rangle\langle\psi(t)|\sqrt{|\psi_0\rangle\langle\psi_0|}}\right)^{\frac{1}{2}} = 0.5 \quad (46)$$

This happens because the initially pure (though superposed) state of (40) becomes a 50:50 mixture of itself and its orthogonal complement due to the random spin-flipping caused by the *x*-basis action of the environment.

We may alternately use as our two-particle system the singlet state:

$$\tfrac{1}{\sqrt{2}}\left(|01\rangle - |10\rangle\right) \quad (47)$$

If we re-express it in the *x*-basis and consider an *x*-basis environment acting upon it, we find that:

$$|\Psi\rangle = \tfrac{1}{\sqrt{2}}\left(|01\rangle - |10\rangle\right)_z = \tfrac{1}{\sqrt{2}}\left(|10\rangle - |01\rangle\right)_x = \tfrac{1}{\sqrt{2}}e^{i\pi}\left(|01\rangle - |10\rangle\right)_x \quad (48)$$

Except for a common phase factor, the state is the same in the *x*-as in the *z*-basis. (More precisely, the *relative state*, alluding to Everett's formulation, is the same.) We may therefore ignore any *x*-basis action of an environment interacting with a system that has this maximum degree of (anti-)symmetry. More generally, no matter what the action basis of an environmental spin, when we rewrite (47) in that basis, we get back the same form. For each of the two-particle entangled "Bell states" (they form a two-particle basis called the "Bell basis"), Table 1 shows what the rewritten system spin looks like when operated upon by an environmental spin acting in a basis defined by the spin vector $\cos\left(\tfrac{\theta}{2}\right)|0\rangle + \sin\left(\tfrac{\theta}{2}\right)|1\rangle$ (without loss of generality we ignore the $\varphi$-direction factor $e^{\pm i\varphi}$ by choosing $\varphi = 0$):

Bell state expressed in the ...

|  ... $z$-basis | ... $\theta$-basis |
|---|---|
| $\beta_{00} = \left( \dfrac{\|00\rangle}{\sqrt{2}} + \dfrac{\|11\rangle}{\sqrt{2}} \right)$ | $\beta_{00}^{(\theta)} = \left( \dfrac{\|00\rangle}{\sqrt{2}} + \dfrac{\|11\rangle}{\sqrt{2}} \right)$ |
| $\beta_{01} = \left( \dfrac{\|01\rangle}{\sqrt{2}} + \dfrac{\|10\rangle}{\sqrt{2}} \right)$ | $\beta_{01}^{(\theta)} = \left( \dfrac{\sin\theta\|00\rangle}{\sqrt{2}} + \dfrac{\cos\theta\|01\rangle}{\sqrt{2}} + \dfrac{\cos\theta\|10\rangle}{\sqrt{2}} - \dfrac{\sin\theta\|11\rangle}{\sqrt{2}} \right)$ |
| $\beta_{10} = \left( \dfrac{\|00\rangle}{\sqrt{2}} - \dfrac{\|11\rangle}{\sqrt{2}} \right)$ | $\beta_{10}^{(\theta)} = \left( \dfrac{\cos\theta\|00\rangle}{\sqrt{2}} - \dfrac{\sin\theta\|01\rangle}{\sqrt{2}} - \dfrac{\sin\theta\|10\rangle}{\sqrt{2}} - \dfrac{\cos\theta\|11\rangle}{\sqrt{2}} \right)$ |
| $\beta_{11} = \left( \dfrac{\|01\rangle}{\sqrt{2}} - \dfrac{\|10\rangle}{\sqrt{2}} \right)$ | $\beta_{11}^{(\theta)} = \left( \dfrac{\|01\rangle}{\sqrt{2}} - \dfrac{\|10\rangle}{\sqrt{2}} \right)$ |

TABLE 1: Basis Transformations for the Bell States

We see that we also get back both $\beta_{11}$ and $\beta_{00}$ unchanged. But $\beta_{00}$ is *completely* decohered by an *x*-basis environment because its two components are affected by the environment in opposite ways. The factors in the diagonal elements of the reduced density matrix contributed by the propagator therefore cancel, whereas those in the off-diagonal elements reinforce. The former elements persist stably, the latter become multiperiodic and vanish. That $\beta_{11}$ remains invariant under a change of basis means that it is 100% invulnerable to decoherence, regardless of the action basis of the environment.

To simulate realistically both decoherence and an instance of a DFS, we have been able to utilize a greatly simplified many-to-one (or -to-two) Hamiltonian. In principal, we could perform a similar transformation on a real environment and work entirely in the resulting basis. In practice we cannot do this: First, real environments are too large to be able to calculate the change of basis matrices; Second, we rarely know the individual coupling constants anyway. For these reasons, the most direct approach to identifying a DFS is in the context of environments that are by nature "collective", meaning that the net environmental interaction with each system spin is the same.

This restriction is not so severe: It comports with the statistical "system-bath" model that is widely applicable anyway. However, a problem does arise when we consider that most systems are affected not only by interactions with the external environment but with the "internal" environment as well [9], that is to say, by interactions among the particles that constitute the system itself. While this is obviously meaningless when the system is a single particle, a DFS requires at least two particles and these may indeed interact. It is precisely their interaction that in the first place may have generated the entangled two-particle state that spans the DFS.

One approach to confirming the existence of a DFS experimentally has been to generate the desired entangled pair via some (possibly only imprecisely understood) interaction but in such a way that the particles immediately separate and the interaction ceases—e.g., the generation of entangled but oppositely-moving photon-pairs via parametric down-conversion [5]. But for systems that require continuous control over a fixed set of entangled particles this solution will be impractical.

Another approach is the diagonalization procedure we used for simulation purposes. If we could diagonalize the system particles only, and if we could tinker with the coefficients of the original system two-particle superposition at will, we can ensure that when the basis is changed, the result is a two-quasi-particle entangled state of the desired form.

However, to diagonalize the system requires that we know the coupling constants between the environmental spins and the systems spins, even if we may ignore those among the environmental spins only. (The only situation where we may ignore the system-environmental coupling is if the system spins form a DFS already—at least initially. Then we would have to deal with their internal interaction only.)

It is instructive to suppose at first that we do have access to the environmental couplings. For two of the four Bell basis states, any tinkering will need to be specific (and may or may not be feasible), namely, for $\beta_{01}$ and $\beta_{10}$, reversing the change of basis expressed in Table 1. But of these two, $\beta_{10}$ is undesirable anyway—it lacks the required symmetry properties in any basis. For the other two Bell states, $\beta_{00}$ and $\beta_{11}$, no tinkering is required anyway: The form of these states is invariant. And of them, $\beta_{00}$ is useless: It decoheres completely regardless of the basis of action. We are left, therefore with finding a basis in which our entangled pair has one of only two desirable Bell-state expressions: either the symmetric triplet state $\beta_{01}$ or the anti-symmetric singlet state $\beta_{11}$. To obtain $\beta_{01}$ we need to create (and design operators that manipulate) the state:

$$\beta_{01}^{(\theta)} = \left( \frac{\sin\theta |00\rangle}{\sqrt{2}} + \frac{\cos\theta |01\rangle}{\sqrt{2}} + \frac{\cos\theta |10\rangle}{\sqrt{2}} - \frac{\sin\theta |11\rangle}{\sqrt{2}} \right) \quad (49)$$

where $\theta$ is defined by the coupling between the two system particles. Ignorance of the environment-system couplings will preclude our doing so. But for $\beta_{11}$, the native and the transformed expressions are identical—and with respect to a DFS it is a superior state anyway.

Therefore, once we have established our system particles in the $\beta_{11} = \frac{1}{\sqrt{2}}(|01\rangle - |10\rangle)$ entangled state, we needn't be concerned that internal interaction among its two constituent particles will decohere it: In principal, it is as resistant to decoherence from its internal environment as from an external one.

Most efforts at preserving coherence aim at isolating a system qubit from the environment This task is challenging because, as the simulations illustrate, decoherence is so efficient that a minute disturbance is sufficient to destroy almost immediately our ability to detect an initial quantum superposition.

Of the successful proof-of-principle implementations of qubit manipulation, a number happen incidentally to have taken advantage of the relatively long lifetimes of states that are rather similar to the anti-symmetric singlet state of two spins—e.g., cavity QED [10], which couples the 0 and 1 number (photon) states of a resonator (in effect a quantum harmonic oscillator) to two excited states of a Rydberg atom. This fact, and the analyses in this paper, suggest that it might be fruitful deliberately to seek physical instances of well-localized quantum states that are from the start in (or that can be placed into) a genuinely anti-symmetric state. Rather than the usual single-particle system, this specific entangled subspace of the larger multiparticle separable Hilbert space—or multi-*quasi-particle* Hilbert space—would then become the standard operational ("computational") basis, and two-(quasi)particle systems in the typical physical substrate for the single qubit.

Theoretical investigations from a very different direction are perhaps leading to the same conclusion. For example, so-called "topological quantum computation" uses as its theoretical qubits a set of mathematical objects that represent distinct vortex-like planar exchanges in position between particles. The statistics associated with these vortices, treated as quasi-particles (quasi-holes, actually), are fractional, not integer. Thus, in one example it requires two additional physical two-state particles to double the size of the state-space, which therefore goes up as $\sim\sqrt{2}^N$, as though each particle had available to it neither 1 nor 2 states but a fractional number of states, i.e. $1 \leq \sqrt{2} \leq 2$ [11]. These quasi-particles are argued to be inherently robust against decoherence because, in effect, the individual qubits are "smeared out" across all the physical particles, holographically (as it is sometimes termed), though a closer analogy is to memories in spin glasses and to Hopfield-type neural networks [12]. Vortices (the quasi-particles) in 2D electron gases are a possible physical example [13, 14]. But the de-localized "smearing" is equivalent to a symmetrization of the basis.

This dependence upon symmetry for protection against decoherence may be analogous to the entangled subspace discussed above [15] where, as well, every two particles contribute not four but two states.

It is a pleasure to acknowledge the contribution of Erich Poppitz of Yale University to many components of this paper, especially those related to the Bogoliubov transformation.